\documentclass[prd,aps,showpacs,notitlepage,superscriptaddress,nofootinbib,preprintnumbers]{revtex4-1}
\usepackage{amsmath}
\usepackage{epsfig}
\def\bea#1\eea{\begin{align}#1\end{align}} 
\newcommand{\nnu}{\nonumber\\}
\newcommand{\bef}{\begin{figure}[htb]\centering}
\newcommand{\eef}{\end{figure}}
\newcommand{\hx}{\hat{x}}
\newcommand{\hz}{\hat{z}}
\newcommand{\pht}{p_{h\perp}}
\newcommand{\bpht}{{\bf p_{h\perp}}}
\newcommand{\xb}{x_B}
\newcommand{\zh}{z_h}

\begin{document}
\preprint{JLAB-THY-14-1947}
\preprint{LA-UR-14-27352}
\title{Next-to-leading order transverse momentum-weighted Sivers asymmetry \\
in semi-inclusive deep inelastic scattering: the role of the three-gluon correlator}

\author{Ling-Yun Dai}
\email{lingyun@jlab.org}
\affiliation{Jefferson Lab, 
                   12000 Jefferson Avenue, 
                   Newport News, VA 23606, USA}

\author{Zhong-Bo Kang}
\email{zkang@lanl.gov}
\affiliation{Theoretical Division, 
                   Los Alamos National Laboratory, 
                   Los Alamos, NM 87545, USA}
                   
\author{Alexei Prokudin}
\email{prokudin@jlab.org}
\affiliation{Division of Science, 
                   Penn State Berks, 
                   Reading, PA 19610, USA}
\affiliation{Jefferson Lab, 
                   12000 Jefferson Avenue, 
                   Newport News, VA 23606, USA}                   
                   
\author{Ivan Vitev}
\email{ivitev@lanl.gov}
\affiliation{Theoretical Division, 
                   Los Alamos National Laboratory, 
                   Los Alamos, NM 87545, USA}
                                      
\begin{abstract}
We study the Sivers asymmetry in semi-inclusive hadron production in deep inelastic scattering. We concentrate on the contribution from the photon-gluon fusion channel at ${\cal O}(\alpha_{em}^2 \alpha_s)$, where three-gluon correlation functions play a major role within the twist-3 collinear factorization formalism. We establish the correspondence between such a formalism with three-gluon correlation functions and the usual transverse momentum dependent (TMD) factorization formalism at moderate hadron transverse momenta. We derive the coefficient functions used in the usual TMD evolution formalism  related to the quark Sivers function expansion in terms of the three-gluon correlation functions. We further perform the next-to-leading order calculation for the transverse-momentum-weighted spin-dependent differential cross section, and identify the off-diagonal contribution from the three-gluon correlation functions to the QCD collinear evolution of the twist-3 Qiu-Sterman function.
\end{abstract}

\date{\today}
\maketitle

%
%

\section{Introduction}
In recent years, the study of single transverse-spin asymmetries (SSAs) has become a forefront of both experimental and theoretical research in QCD and hadron physics. With extensive experimentation underway and major theoretical advances, we have begun to obtain a deeper understanding of the nucleon structure and the partons' transverse motion. A lot of progress was made in understanding  the underlying QCD mechanisms that generate these asymmetries. The  transverse momentum dependent (TMD) factorization scheme~\cite{Ji:2004wu,Ji:2004xq,Collins:2011zzd} and the twist-3 collinear factorization approach~\cite{Efremov:1981sh,Efremov:1984ip,Qiu:1991pp} were studied theoretically and applied phenomenologically to describe the  SSAs in various processes, including Drell-Yan, semi-inclusive deep inelastic scattering (SIDIS), $e^+e^-$ annihilation, and hadron and jet production in $pp$ scattering. These two mechanisms were shown to be closely related and provide a unified picture for SSAs~\cite{Ji:2006ub,Bacchetta:2008xw}.

SIDIS is one of the key experimental tools to study the spin asymmetries and the associated nucleon structure. 
A particular twist-3 quark-gluon correlation function, often called Qiu-Sterman function \cite{Qiu:1991pp}, plays a crucial role in generating non-zero SSA and is related to the quark Sivers function \cite{Sivers:1989cc,Boer:1997nt}. SSAs enabled by the Sivers function were extensively studied experimentally in the SIDIS process by HERMES \cite{Airapetian:2009ae}, COMPASS \cite{Alekseev:2008aa,Adolph:2013stb,Adolph:2014zba} and JLab \cite{Qian:2011py}. It was discovered theoretically that the Sivers function should change sign when measured in the Drell-Yan process with respect to the SIDIS process \cite{Brodsky:2002cx,Collins:2002kn,Kang:2009bp} and a number of experiments, including COMPASS, RHIC experiments, and Fermilab experiments, are planned to test this prediction experimentally.
Knowledge of the evolution of the Sivers function \cite{Kang:2011mr,Aybat:2011ge,Echevarria:2012pw,Sun:2013hua,Echevarria:2014xaa} (and Qiu-Sterman function \cite{Kang:2008ey,Zhou:2008mz,Vogelsang:2009pj,Braun:2009mi,Kang:2012em,Schafer:2012ra,Kang:2010xv,Schafer:2013opa}) with the hard scale is very important for accurate phenomenological applications and, eventually, for precise extraction of these functions. Next-to-leading order (NLO) corrections involving the Qiu-Sterman function were calculated for Drell-Yan \cite{Vogelsang:2009pj} and SIDIS~\cite{Kang:2012ns}.

Special three-gluon correlation functions~\cite{Ji:1992eu,Eguchi:2006qz,Eguchi:2006mc,Beppu:2010qn,Kang:2008ey,Kang:2008qh,Kang:2008ih} become
 relevant at NLO  ($ {\cal O}(\alpha_{em}^2 \alpha_s)$) and can be studied experimentally via open charm production in SIDIS \cite{Kang:2008ih,Beppu:2010qn}. The purpose of our current paper is to study the role of three-gluon correlation functions in SIDIS, and their  connection to the quark Sivers function. Concentrating on the photon-gluon fusion channel, we first calculate the contributions of the three-gluon correlation functions to the transverse spin-dependent differential cross section within the twist-3 collinear factorization formalism. We then demonstrate that our result can be matched onto the TMD factorization formalism at  moderate hadron transverse momenta, and that we can extend the unification of the two mechanisms to the case involving three-gluon correlation functions. We also derive the  coefficient functions widely used in the TMD evolution formalism. This is achieved by expanding the quark Sivers function (in the Fourier transformed $b$-space) in terms of a convolution of the coefficient functions and the three-gluon correlation functions. 

In the second part of our paper, we study the NLO perturbative QCD corrections to the transverse momentum-weighted spin-dependent SIDIS cross section. Our primary focus is again on the contributions of the three-gluon correlation functions. By analyzing the collinear divergence structure, we identify the evolution kernel for the Qiu-Sterman function that includes the off-diagonal contribution from the three-gluon correlators. The hard coefficient function is evaluated at one-loop order.

%
%
\section{Sivers asymmetry from three-gluon correlation functions}
\label{SecII}

In this section we first study  the contribution of the three-gluon correlation functions to the Sivers asymmetry in SIDIS  within the twist-3 collinear factorization formalism. We establish the correspondence between such a formalism and the usual TMD factorization formalism at  moderate hadron transverse momenta to be defined below. Coincidently, we derive the coefficient functions $C_{q\leftarrow g}$ by expanding the quark Sivers function in the conjugate Fourier $b$-space in terms of the three-gluon correlation functions. Such coefficient functions are a key ingredient of the usual TMD evolution formalism.

%
%
\subsection{Three-gluon correlation functions}
To define the twist-3 three-parton correlation functions, we consider a nucleon of momentum $p^\mu=p^+\bar n^\mu$, with $\bar n^\mu = [1^+,0^-,\bf{0_\perp}]$ expressed in light-cone coordinates, where we write any four-vector $v^\mu = [v^+, v^-, {\bf v_\perp}]$ with $v^+ = \frac{1}{\sqrt{2}}(v^0 + v^z)$ and $v^- = \frac{1}{\sqrt{2}}(v^0 - v^z)$. We also define a conjugated light-like vector $n^\mu = [0^+,1^-,\bf{0_\perp}]$, which obeys $n \cdot  \bar n = 1$, $n^2 = 0$, and $\bar n^2 = 0$. The widely studied twist-3 quark-gluon correlation function $T_{q,F}(x_1, x_2)$ (the so-called ``Qiu-Sterman'' function) is defined as follows~\cite{Kang:2011hk}:
\bea
M_{F,aij}^{\alpha}(x_1, x_2) &= g_s \int\frac{dy_1^-dy_2^-}{2\pi}  e^{i x_1 p^+ y_1^-}  e^{i (x_2-x_1) p^+ y_2^-} 
\langle p\, s|\bar{\psi}_j(0) F_a^{\alpha+}(y_2^-) \psi_i(y_1^-)  |p\, s\rangle
\nnu&=
\frac{1}{2} \left[T_{q,F}(x_1, x_2) \gamma\cdot \bar n \epsilon^{\alpha n \bar n s} \frac{2}{N_c^2-1} \left(t^a\right)_{ij} + \cdots \right],
\eea
where $|p\,s\rangle$ represents the nucleon wave-function with $p$ the momentum of the nucleon given above, and spin vector $s^\mu = (0, 0, {\mathbf s_\perp})$. $g_s$ is the strong coupling, $\psi_{i,j}$ are the quark fields with color indices $i,~j$ in the fundamental representation of the color $SU(N_c)$ group. We have $N_c=3$ the number of the colors, $t^a$ is the standard generating matrix of the $SU(N_c)$ group. $F_a^{\alpha+} = F_a^{\alpha\beta}n_\beta$ with $F_a^{\alpha\beta}$ the standard gluon field strength. $x_{1,2}$ are the momentum fractions carried by the quarks (represented by $\psi_{i}$ and $\psi_j$, respectively), and $(x_2-x_1)$ will be the momentum fraction for the gluon following the momentum conservation. Note that $\epsilon^{\alpha n \bar n s} = \epsilon^{\alpha \beta \mu \nu} n_{\beta} \bar n_{\mu} s_{\nu}$ with $\epsilon^{\alpha \beta \mu \nu}$ the Levi-Civita tensor, and we use the convention $\epsilon^{0123} = +1$ here.

Classification of three-gluon correlation functions was first considered in \cite{Ji:1992eu}. These functions have been studied in the context of open charm production in \cite{Kang:2008qh,Kang:2008ih,Eguchi:2006qz,Eguchi:2006mc,Beppu:2010qn}. Generically, three-gluon correlation functions can be constructed as combinations of the  gauge invariant correlation functions $\langle i f^{abc} F_a^{\alpha +} F_b^{\beta +}F_c^{\gamma +}\rangle$ and $\langle d^{abc} F_a^{\alpha +} F_b^{\beta +}F_c^{\gamma +}\rangle$, where $f^{abc}$ and $d^{abc} $ are the anti-symmetric and symmetric structure constants of the $SU(N_c)$ color group. With slightly different normalization from 
Ref.~\cite{Beppu:2010qn}, we define the following three-gluon correlation function:
\bea
M^{\alpha\beta\gamma}_{F,abc}(x_1,x_2) &= g_s \int\frac{dy_1^-dy_2^-}{2\pi}  e^{i x_1 p^+ y_1^-}  e^{i (x_2-x_1) p^+ y_2^-} \frac{1}{p^+} \langle p\, s| F_b^{\beta +} (0)F_c^{\gamma +} (y_1^-)F_a^{\alpha +} (y_2^-)  |p\, s\rangle  \nonumber \\
&=  ({\cal C}^{(d)}_g)^{abc} O^{\alpha\beta\gamma}(x_1,x_2)  -  ({\cal C}^{(f)}_g)^{abc} N^{\alpha\beta\gamma}(x_1,x_2)\; ,
\label{eq:MF} 
\eea  
where the gluonic color projection operators $({\cal C}^{(d)}_g)^{abc}$ and $({\cal C}^{(f)}_g)^{abc}$ are given by
\bea
({\cal C}^{(d)}_g)^{abc} &= \frac{N_c}{(N_c^2-1)(N_c^2-4)} d^{abc},
\\
({\cal C}^{(f)}_g)^{abc} &= \frac{i}{N_c(N_c^2-1)} f^{abc}.
\eea
The functions  $O^{\alpha\beta\gamma}(x_1,x_2)$ and  $N^{\alpha\beta\gamma}(x_1,x_2)$ correspond to symmetric and anti-symmetric combinations of gluon field-strength tensors and read \cite{Beppu:2010qn}
\bea
O^{\alpha\beta\gamma}(x_1,x_2) &=   g_s \int\frac{dy_1^-dy_2^-}{2\pi}  e^{i x_1 p^+ y_1^-}  e^{i (x_2-x_1) p^+ y_2^-} \frac{1}{p^+} \langle p\, s| d^{bca} F_b^{\beta +} (0)F_c^{\gamma +} (y_1^-)F_a^{\alpha n} (y_2^-)  |p\, s\rangle 
\nnu
&= \frac{1}{2} \left[O(x_1,x_2) g_\perp^{\alpha\beta} \epsilon^{\gamma n \bar n  s} + O (x_2,x_2-x_1) g_\perp^{\beta\gamma} \epsilon^{\alpha n \bar n  s} + O (x_1,x_1-x_2) g_\perp^{\gamma\alpha} \epsilon^{\beta n \bar n s}\right] \; ,
\label{eq:o}
\\
N^{\alpha\beta\gamma}(x_1,x_2) &=   g_s \int\frac{dy_1^-dy_2^-}{2\pi}  e^{i x_1 p^+ y_1^-}  e^{i (x_2-x_1) p^+ y_2^-} \frac{1}{p^+} \langle p\, s| i f^{bca} F_b^{\beta +} (0)F_c^{\gamma +} (y_1^-)F_a^{\alpha n} (y_2^-)  |p\, s\rangle 
\nnu
&= \frac{1}{2} \left[N (x_1,x_2) g_\perp^{\alpha\beta} \epsilon^{\gamma n \bar n s} - N (x_2,x_2-x_1) g_\perp^{\beta\gamma} \epsilon^{\alpha n \bar n s} - N (x_1,x_1-x_2) g_\perp^{\gamma\alpha} \epsilon^{\beta n \bar n s}\right] \; ,
\label{eq:n}
\eea  
where $ g_\perp^{\alpha\beta} = -  g^{\alpha\beta}+\bar n^\alpha n^\beta +\bar n^\beta n^\alpha$. Our definitions are related to those of Refs.~\cite{Eguchi:2006qz,Eguchi:2006mc,Beppu:2010qn} by Koike et.al. as follows:
\bea
O (x_1,x_2) &=  8 \pi M  \, O(x_1,x_2)|_{\rm Koike} \; , 
\nnu
N (x_1,x_2) &=  8 \pi M \, N(x_1,x_2)|_{\rm Koike} \; ,
\eea
with $M$ being the nucleon mass. 

%
%
\subsection{Spin-dependent cross section for SIDIS: three-gluon correlation functions}
\label{cross section}
We now consider the contribution of the three-gluon correlation functions to the Sivers asymmetry for the SIDIS process, $e(\ell)+p(p, {\bf s_\perp})\to e(\ell')+h(p_h)+X$. Here ${\bf s_\perp}$ is the transverse spin vector of the incoming nucleon with momentum $p$, whereas $\ell$ and $\ell'$ are the momenta of the lepton before and after the collision. $h$ represents the observed final-state hadron with momentum $p_h$, and the exchanged virtual photon has momentum $q=\ell - \ell'$ with the invariant mass $Q^2 = -q^2$.  We will work in the so-called {\it hadron frame}~\cite{Meng:1991da,Koike:2003zc,Kang:2008qh,Beppu:2010qn}, where both the virtual photon $q$ and the incoming polarized nucleon $p$ have only the $z$-component, i.e., vanishing transverse momentum. In this frame, the final observed hadron has transverse momentum, which will be denoted as ${\bf p_{h\perp}}$ below with its magnitude $\pht = |\bpht|$. 

This process has already been studied in \cite{Beppu:2010qn}, for open charm production. We will reproduce the result here. However, the purpose of our calculation is quite different. What we investigate here is  the connection between the twist-3 formalism and the TMD factorization approach, in particular, for the three-gluon correlation functions. In the course of such a study, we will further derive the contribution of three-gluon correlation functions to the evolution of Qiu-Sterman function $T_{q,F}(x, x)$. Finally, we will derive the so-called coefficient functions which are widely used in the TMD evolution formalism. Except for the very first result available in \cite{Beppu:2010qn}, all other calculations (matching, evolution, and coefficient functions) are performed only in the current paper. Because of the different goal in our calculations, we will study {\it light} hadron production, i.e., we consider the mass of the hadron is much smaller than the hard scale $p_h^2 = m_h^2 \ll Q^2$.

The differential cross section that includes the  Sivers effect, i.e. the $\sin(\phi_h - \phi_s)$ module, can be written as follows \cite{Bacchetta:2006tn,Kang:2012xf,Kang:2012ns}:
\bea
\frac{d\sigma_{\rm Sivers}}{dx_B dy dz_h d^2\bpht} = \sigma_0 \left [ F_{UU} + \sin(\phi_h-\phi_s) 
F_{UT}^{\sin(\phi_h-\phi_s)}\right],
\eea
where $F_{UU}$ and $F_{UT}^{\sin(\phi_h-\phi_s)}$ are the spin-averaged and transverse spin-dependent structure functions, respectively, and $\phi_h, \, \phi_s$ are the azimuthal angles of the final-state hadron transverse momentum $\bpht$ and the proton spin ${\mathbf s_\perp}$ relative to the scattering plane of the lepton. $\sigma_0$ is given by
\bea
\sigma_0 = \frac{2 \pi \alpha_{em}^2}{Q^2}\frac{1+(1-y)^2}{y},
\eea
and $x_B$, $y$, and $z_h$ are the standard SIDIS kinematic variables,
\bea
S=(p+\ell)^2, 
\qquad
\xb=\frac{Q^2}{2p\cdot q}, 
\qquad
y=\frac{p\cdot q}{p\cdot \ell} = \frac{Q^2}{x_B S}, 
\qquad
\zh=\frac{p\cdot p_h}{p\cdot q}.
\eea

The transverse spin-dependent differential cross section $d\Delta\sigma/{dx_B dy dz_h d^2\pht}  \equiv \sigma_0  \sin(\phi_h-\phi_s) F_{UT}^{\sin(\phi_h-\phi_s)}$, which is the main focus of this section. It can be written as~\cite{Beppu:2010qn}
 \bea
\frac{d \Delta\sigma}{dx_B dy dz_hd^2\bpht}= \frac{\alpha_{em}^2 y}{32 \pi^3 Q^4 z_h} L^{\mu\nu} W_{\mu\nu} (p,q, p_h) \; ,
\label{eq:crs}
\eea
where $L^{\mu\nu} = 2(\ell_\mu \ell'_\nu + \ell_\nu \ell'_\mu) - Q^2 g_{\mu\nu}$ is the leptonic tensor and $W^{\mu\nu}$ is the hadronic tensor. The hadronic tensor $W^{\mu\nu}$ is related to the partonic tensor $w^{\mu\nu}$ by
\bea
W^{\mu\nu}(p,q,p_h) = \int \frac{d z}{z^2} D_{h/q}(z) w^{\mu\nu}(p,q,p_c) \; ,
\eea 
where $D_{h/q}(z)$ is the fragmentation function of a quark $q$ into a hadron $h$, and the parton momentum $p_c^\mu = p_h^\mu/z$. In the following 
(and throughout the paper) we will only consider the so-called metric contribution \cite{Graudenz:1994dq,Daleo:2004pn,Kang:2014ela,Kang:2013raa}, i.e. we contract $w^{\mu\nu}$ with $-g_{\mu\nu}$ and write 
$w=\left[-g_{\mu\nu} w^{\mu\nu}\right]$ below. Within the collinear factorization formalism, the transverse spin-dependent cross section is a twist-3 effect. To extract this effect, one has to perform a collinear expansion around a vanishing parton $k_\perp$. For three-gluon correlation functions, the contribution can be written as \cite{Beppu:2010qn}
\bea
w(p,q,p_c) = \int \frac{d x_1}{x_1}\frac{d x_2}{x_2} \frac{\partial }{\partial k^\lambda_\perp} \Big[H_{\rho\delta\sigma}^{abc}(p,q,p_c,k_\perp) p^\delta\Big]_{k_\perp\to 0} \, \omega^{\rho}_{\alpha} \omega^{\sigma}_{\beta}
\omega^{\lambda}_{\gamma} \, M^{\alpha\beta\gamma}_{F,abc} (x_1,x_2) \, ,
\label{eq:w}
\eea
with $\omega^{\mu}_{\nu} = \delta^{\mu}_{\nu} - \bar n^{\mu} n_{\nu}$. A generic diagram to calculate the photon-gluon hard-part function $H_{\rho\delta\sigma}^{abc}$ is sketched in Fig.~\ref{fig:amplitude}.

\bef    
\psfig{file=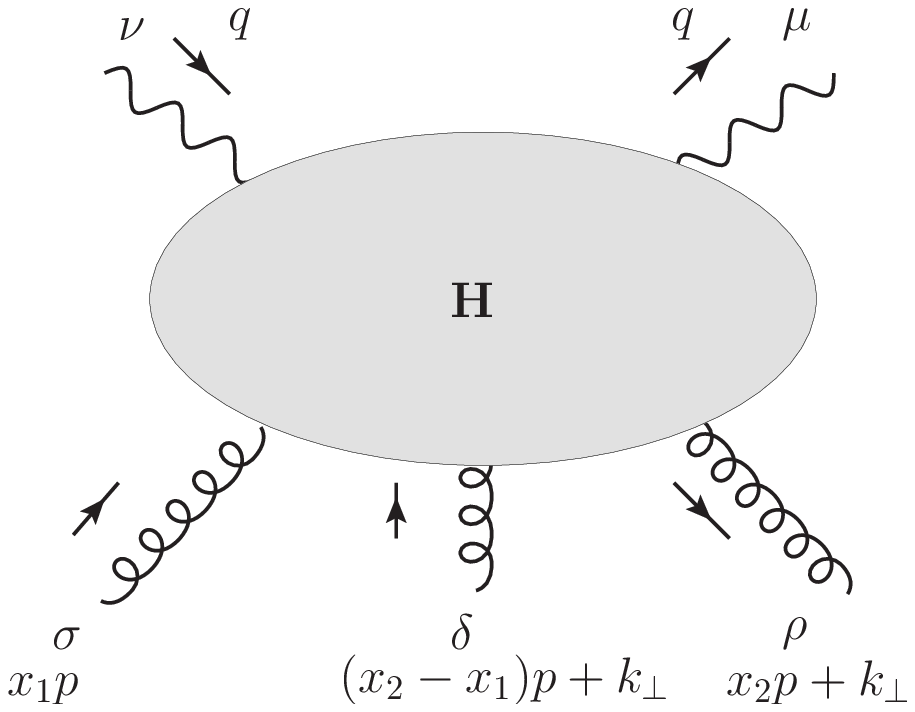, scale=0.45}
\caption{Generic diagram that is used to calculate the hard-part function $H_{\rho\delta\sigma}^{abc}$.}
\label{fig:amplitude}
\eef

Using Eqs.~\eqref{eq:MF}, \eqref{eq:o}, and \eqref{eq:n}, we can rewrite Eq.~\eqref{eq:w} as
\bea
w(p,q,p_c) = \int \frac{d x_1}{x_1}\frac{d x_2}{x_2} \frac{\partial }{\partial k^\lambda_\perp} \Big[H_{\rho\delta\sigma}^{abc}(p,q,p_c,k_\perp) p^\delta\Big]_{k_\perp\to 0} \,  F_{NO}^{\rho\sigma\lambda}(x_1,x_2) \, ,
\eea
where $F_{NO}^{\rho\sigma\lambda}(x_1,x_2)$ represents
\bea
F_{NO}^{\rho\sigma\lambda}(x_1,x_2) =& ({\cal C}^{(d)}_g)^{abc} \left( O (x_1,x_2) g_\perp^{\rho\sigma} \epsilon^{\lambda n \bar n s} + O (x_2,x_2-x_1) g_\perp^{\sigma\lambda} \epsilon^{\rho n \bar n s} + O (x_1,x_1-x_2) g_\perp^{\lambda\rho} \epsilon^{\sigma n \bar n s}\,\right) 
\nnu
&
- ({\cal C}^{(f)}_g)^{abc} \left( N (x_1,x_2) g_\perp^{\rho\sigma} \epsilon^{\lambda n \bar n s} - N (x_2,x_2-x_1) g_\perp^{\sigma\lambda} \epsilon^{\rho n \bar n s} - N (x_1,x_1-x_2) g_\perp^{\lambda\rho} \epsilon^{\sigma n \bar n s}\,\right) \, .
\label{eq:FNO}
\eea
The relevant Feynman diagrams for the transverse momentum dependent differential cross section $d\Delta\sigma/{dx_B dy dz_h d^2\pht}$ at leading order (LO) are listed in Fig.~\ref{fig:soft-pole}. The technique to extract twist-3 contributions is well explained in the literature. The idea is that the so-called ``pole-propagators'' and the on-mass-shell condition for the unobserved parton in the final-state lead to kinematic $\delta$-functions, which can be used to integrate out the parton momentum fractions $x_1$ and $x_2$. These parton momentum fractions $x_{1,2}$ generally depend on $k_\perp$, and, thus, are expanded with respect to $k_\perp$. After some algebraic manipulation we have the following ``master formula'':
\bea
w(p,q,p_c) =& (v_1 -v_2)_\lambda \frac{1}{x^2} \left( \frac{d F_{NO}^{\rho\sigma\lambda}(x,x)}{d x} - \frac{2 F_{NO}^{\rho\sigma\lambda}(x,x)}{x}\right) H^L_{\rho\sigma}(x, x, 0 )  +  \frac{F_{NO}^{\rho\sigma\lambda}(x,x)}{x^2}  
\nnu
& \times   
\lim_{k_\perp \rightarrow 0} \frac{\partial}{\partial k_\perp^\lambda}\left[H^L_{\rho\sigma}(x+ (v_2 -v_1)\cdot k_\perp, x+ v_2\cdot k_\perp, k_\perp ) - H^R_{\rho\sigma}(x, x+ v_1\cdot k_\perp, k_\perp )\right]   ,
\label{eq:w1}
\eea
where $H^{L}_{\rho\sigma}$ are remainders of the hard parts $[H^{}_{\rho\delta\sigma} p ^\delta]$ given in Fig.~\ref{fig:soft-pole}, while $H^{R}_{\rho\sigma}$ are the mirror diagrams where the middle gluon is to the right of the unitary cut.  The two four-vectors $v_1$ and $v_2$ are given by
\bea
v_1^\mu = -\frac{2 x}{\hat u } p_c^\mu,
\qquad
v_2^\mu = -\frac{2 x}{\hat t } p_c^\mu,
\eea
with the partonic Mandelstam variables 
\bea
\hat s = (x p + q)^2 \, , 
\qquad
\hat t = (q - p_c)^2 \, ,
\qquad
\hat u = (x p - p_c)^2 \, .
\eea
 
\bef
\psfig{file=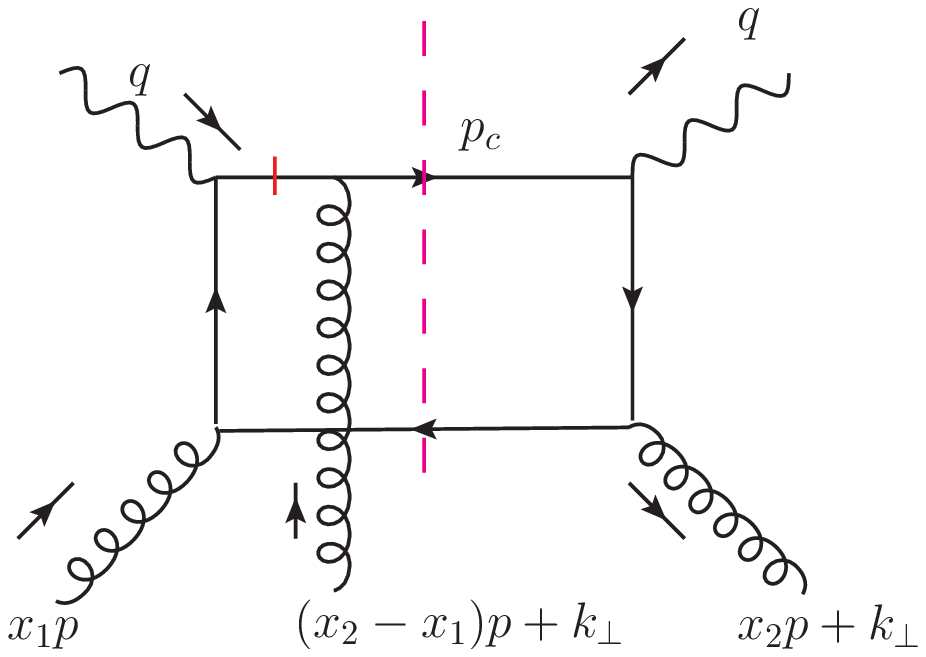,  scale=0.45}
\hskip 0.1in
\psfig{file=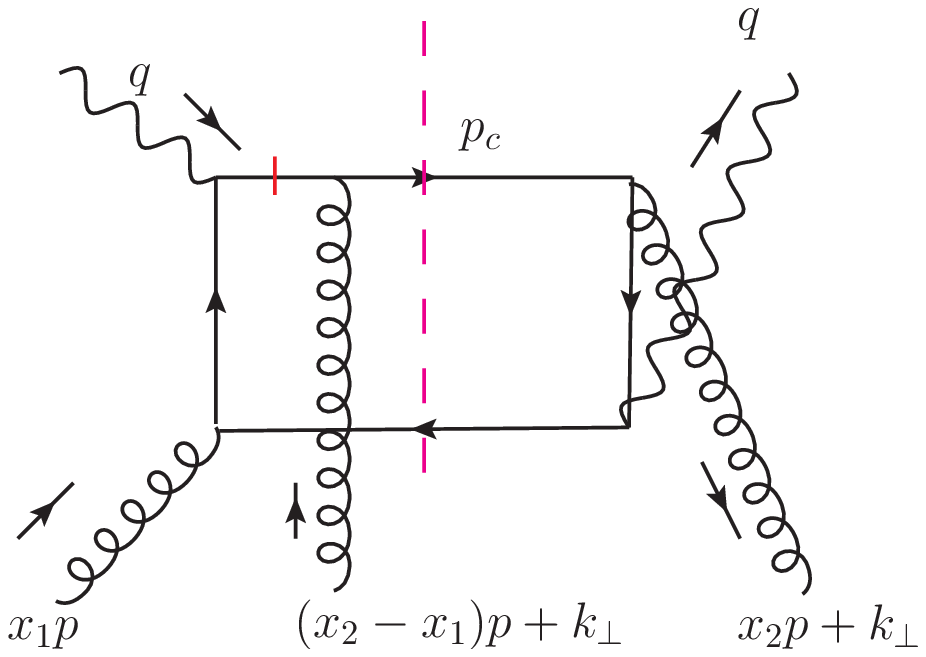, scale=0.45}
\\
\vskip 0.1in
\psfig{file=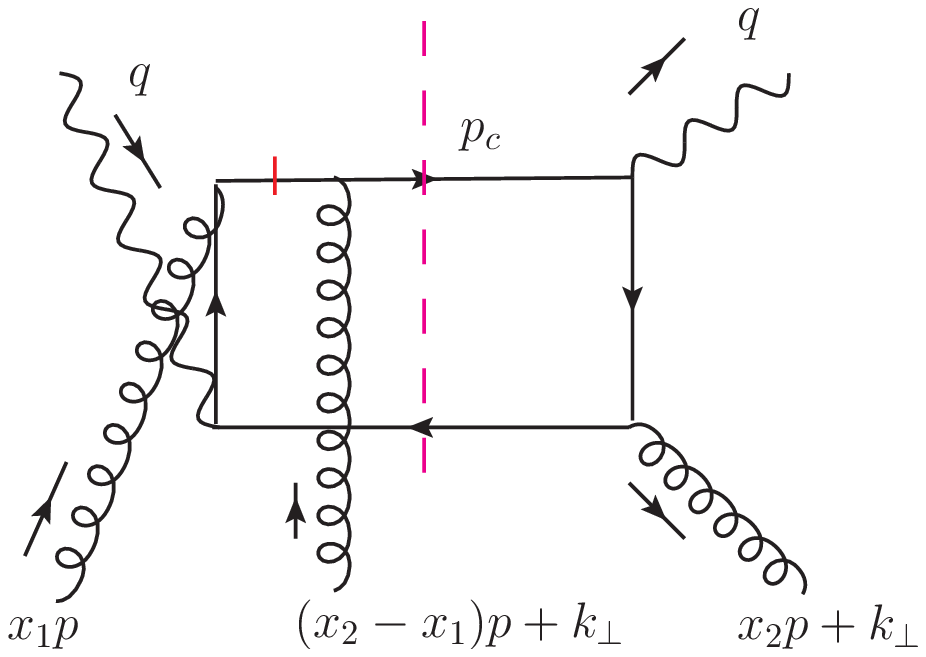, scale=0.45}
\hskip 0.1in
\psfig{file=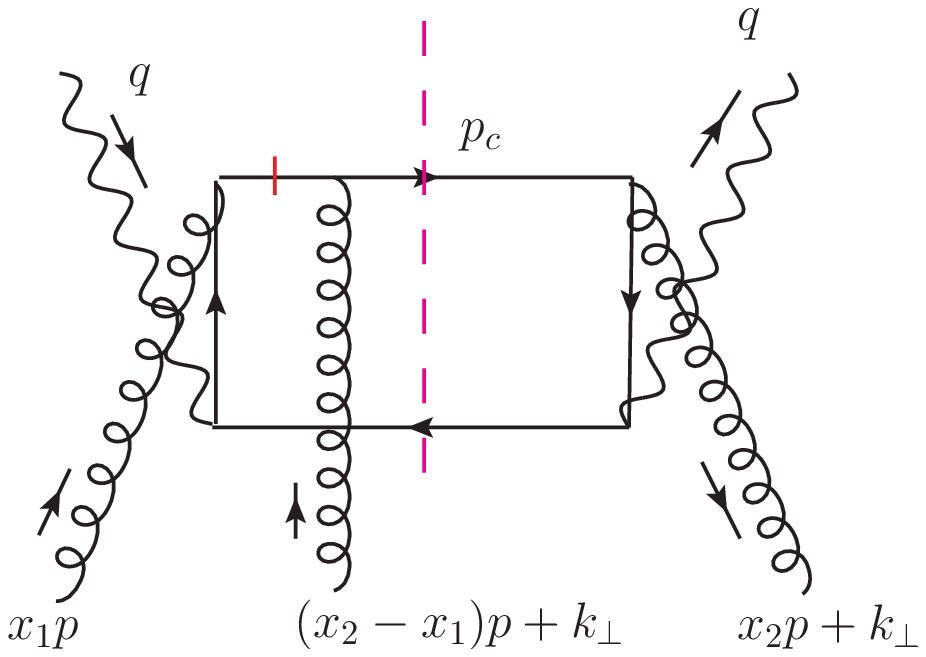, scale=0.45}
\caption{Feynman diagrams that enter the calculation of the photon-gluon fusion hard part. The mirror diagrams, where middle gluon is to the right of the unitary cut, also contribute and are included in our final result.}
\label{fig:soft-pole}
\eef

The final result for the transverse spin-dependent differential cross section is given by
\bea
\frac{d\Delta\sigma}{d\xb dy dz_h d^2\bpht} =&  \sigma_0 \left(\epsilon^{\alpha\beta} s_\perp^\alpha 
\pht^\beta\right) \sum_q e_q^2 \left(\frac{1}{4}\right) \frac{\alpha_s}{2\pi^2}
\int_{x_B}^1\frac{dx}{x}\int_{z_h}^1\frac{dz}{z} D_{h/q}(z) 
\frac{1}{zQ^2}\delta\left(\pht^2 - z_h^2Q^2\left(\frac{1}{\hat x} - 1\right)\left(\frac{1}{\hat z} - 1\right)\right)
\nonumber\\
&\times\left\{\left[\left(\frac{d O(x,x)}{d x}-\frac{2 O(x,x)}{x}\right)H_1 + 
                  \left(\frac{d O(x,0)}{d x}-\frac{2 O(x,0)}{x}\right)H_2 + 
                  \frac{O(x,x)}{x}H_3+\frac{O(x,0)}{x}H_4\right]  \right.   
\nonumber\\                   
&+\left.\left[\left(\frac{d N(x,x)}{d x}-\frac{2 N(x,x)}{x}\right)H_1 - 
              \left(\frac{d N(x,0)}{d x}-\frac{2 N(x,0)}{x}\right)H_2
              +\frac{N(x,x)}{x}H_3-\frac{N(x,0)}{x}H_4\right]\right\}  \;.
\label{result-4}              
\eea
where $\epsilon^{\alpha\beta}$ is a two-dimensional anti-symmetric tensor with $\epsilon^{12}=1$, and thus $\epsilon^{\alpha\beta} s_{\perp}^\alpha \pht^{\beta} = \pht \sin(\phi_h-\phi_s)$. Note that the integration limit for $x$ and $z$ are the standard ones, since we are considering  {\it light} hadron production, as we have emphasized in the beginning of this subsection. Thus, there are no specific restrictions on the integration limit for $x$ and $z$ as pointed out in~\cite{Beppu:2010qn}, which only apply when the mass of the hadron is important. The hard-part functions $H_{i=1,2,3,4}$ have the following expressions:
\bea
H_1&=\frac{\hx\left[2\hx^2-2\hx+(1-2\hz+2\hz^2)\right]}{\hz^2(1-\hz)^2}\;,
\\
H_2&=\frac{\hx\left[4\hx^2-4\hx+(1-2\hz+2\hz^2)\right]}{\hz^2(1-\hz)^2}\;,
\\
H_3&=\frac{2\hx^2\left(1-2\hx\right)}{\hz^2(1-\hz)^2}\;,
\\
H_4&=\frac{2\hx^2\left(1-4\hx\right)}{\hz^2(1-\hz)^2}\;,
\eea
with $\hx = x_B/x$ and $\hz=z_h/z$. Our results are consistent with those in \cite{Beppu:2010qn} \footnote{One simply realizes that $\left[-g^{\mu\nu}w_{\mu\nu}\right] = \left(2\widetilde{V}_1^{\mu\nu} - 3\widetilde{V}_2^{\mu\nu}\right)w_{\mu\nu}$ in \cite{Beppu:2010qn}.}.

%
%
\subsection{Matching onto the TMD factorization formalism}
It has been demonstrated that the collinear twist-3 factorization formalism and the TMD factorization formalism are consistent with each other (so-called ``matching'') for  moderate hadron transverse momenta, i.e., in the kinematic region $\Lambda_{\rm QCD} \ll \pht \ll Q$, see e.g., \cite{Ji:2006ub,Ji:2006vf,Ji:2006br,Koike:2007dg,Bacchetta:2008xw,Yuan:2009dw,Boer:2010ya}. However, in these earlier studies the matching was only demonstrated/shown for {\it the quark-gluon correlation function}  $T_{q,F}(x_1,x_2)$.  In this paper we generalize the known correspondence to include the three-gluon correlation functions for the first time. To demonstrate such a connection, we first study the limit of the transverse spin-dependent cross section in Eq.~\eqref{result-4} derived from the collinear twist-3 factorization formalism when $\pht\ll Q$. Using \cite{Ji:2006br}
\bea
\left.\delta\left(\pht^2 - z_h^2Q^2\left(\frac{1}{\hat x} - 1\right)\left(\frac{1}{\hat z} - 1\right)\right)\right|_{\pht \ll Q} = \frac{(1-\hx)(1-\hz)}{\pht^2}  \left[ \frac{\delta(1-\hx)}{(1-\hz)_+} + \frac{\delta(1-\hz)}{(1-\hx)_+}  + \delta(1-\hx)\delta(1-\hz)\ln\left(\frac{z_h^2Q^2}{\pht^2}\right) \right],
\eea
we find that
\bea
\left.\frac{d\Delta\sigma}{d\xb dy dz_h d^2\bpht}\right|_{\pht \ll Q} =& - z_h \sigma_0 \left(\epsilon^{\alpha\beta} s_\perp^\alpha \pht^\beta \right) \frac{1}{\left(\pht^2\right)^2} \sum_q e_q^2
\frac{\alpha_s}{2\pi^2} \int \frac{dz}{z} D_{h/q}(z) \delta(1-\hz)
\nnu
&\times \int \frac{dx}{x^2} P_{q\leftarrow g}(\hx) \left( \frac{1}{2}\right) \left[ O(x,x) + O(x,0) + N(x,x) - N(x,0)\right],
\label{eq:twist-3}
\eea
where $P_{q\leftarrow g}$ is the usual gluon-to-quark splitting kernel
\bea
P_{q\leftarrow g}(\hx) = T_R \left[\hx^2 + (1-\hx)^2\right],
\eea
with the color factor $T_R = \frac{1}{2}$. It is instructive to point out that to arrive at the final result in Eq.~\eqref{eq:twist-3} we have used  integration by parts in Eq.~\eqref{result-4} to convert all the derivative terms to  non-derivative terms, as well as the fact that $O(x,x), O(x,0), N(x,x), N(x,0)$ vanish when parton momentum fraction $x\to 1$.

On the other hand, the TMD factorization formalism \cite{Ji:2004wu,Ji:2004xq,Collins:2011zzd} for the SIDIS process gives 
\bea
\frac{d\Delta\sigma}{d\xb dy dz_h d^2\bpht} =& \sigma_0 \sum_q e_q^2 \int d^2{\bf k_\perp} d^2{\bf p_\perp} d^2\boldsymbol{\lambda_\perp} \delta^2\left(z_h {\bf k_\perp} + {\bf p_\perp} + \boldsymbol{\lambda_\perp} - {\bf p_{h\perp}} \right) 
\nnu
&\times
\frac{\epsilon^{\alpha\beta} s_\perp^\alpha k_\perp^\beta}{M}f_{1T}^{\perp q}(x_B, k_\perp^2) D_{h/q}(z_h, p_\perp^2) S(\boldsymbol{\lambda_\perp}) H(Q^2),
\label{eq:TMD-fac}
\eea
where $f_{1T}^{\perp q}(x_B, k_\perp^2)$ is the quark Sivers function, $D_{h/q}(z_h, p_\perp^2)$ is the transverse momentum dependent fragmentation function, $S(\boldsymbol{\lambda_\perp})$ and $H(Q^2)$ denotes the soft and hard factors, respectively. Note that here both $f_{1T}^{\perp q}(x_B, k_\perp^2)$ and $D_{h/q}(z_h, p_\perp^2)$ are the so-called unsubtracted TMD functions~\cite{Collins:2011zzd}. To make contact with the result from the collinear twist-3 formalism in Eq.~\eqref{eq:twist-3}, we need to compute the perturbative tail of the various factors in the TMD formalism in Eq.~\eqref{eq:TMD-fac}. In particular we need the expansion of the quark Sivers function $f_{1T}^{\perp q}(x_B, k_\perp^2)$ in terms of the three-gluon correlation functions when $k_\perp \gg \Lambda_{\rm QCD}$. This is usually referred to as the so-called {\it off-diagonal} term, where the quark Sivers function receives contributions from the three-gluon correlation functions (quark from gluon), as opposed to the known {\it diagonal} term in \cite{Ji:2006ub,Ji:2006vf,Ji:2006br,Koike:2007dg,Bacchetta:2008xw,Yuan:2009dw,Boer:2010ya}, where the quark Sivers function receives contributions from the quark-gluon correlation function $T_{q,F}(x_1, x_2)$ (quark from quark).

The relevant Feynman diagrams to compute the quark Sivers function $f_{1T}^{\perp q}(x_B, k_\perp^2)$ in terms of the three-gluon correlation functions are shown in Fig.~\ref{sivers-3g}. This is the  forward cut scattering diagram, where the left side of the cut (the magenta dashed line) is the amplitude and the right side of the cut is the conjugate to the amplitude. The upper part of this diagram represents the quark Sivers function with the momentum $k$ for the quark, where $k^+ = x_B p^+$ with $p$ the momentum of the nucleon. Note that the nucleon is represented by the grey blob in the bottom of the diagram. The double line represents the gauge link (eikonal line) in the definition of the quark Sivers function. In the middle part of the diagram, we have three-gluon correlation functions in the nucleon (as represented by three gluons coming out of the nucleon). In other words, such a diagram just represents the contribution of three-gluon correlation functions to the quark Sivers function, which is very similar to those contributions of the quark-gluon correlation function $T_{q,F}(x_1, x_2)$ to the quark Sivers functions, see, e.g., Fig.~9 of~\cite{Ji:2006vf}\footnote{see also the similar Feynman diagram Fig.~8(e) for the collinear gluon distribution contribution to the unpolarized quark TMD~\cite{Ji:2006vf}.}. To obtain the final result, we have to perform the same collinear expansion  as in  Eqs.~\eqref{eq:w} and \eqref{eq:w1}, and the result can be written as the following form
\bea
\frac{1}{M}f_{1T}^{\perp q}(x_B, k_\perp^2) = -\frac{\alpha_s}{2\pi^2} \frac{1}{\left(k_\perp^2\right)^2} \int_{x_B}^1 \frac{dx}{x^2}  P_{q\leftarrow g}(\hat x) \left(\frac{1}{2}\right)\left[O(x,x) + O(x,0) + N(x,x) - N(x,0)\right],
\label{eq:sivers-expand}
\eea
where to arrive at the above result we have again used integration by parts to convert all derivative terms to non-derivative terms. 
\bef
\psfig{file=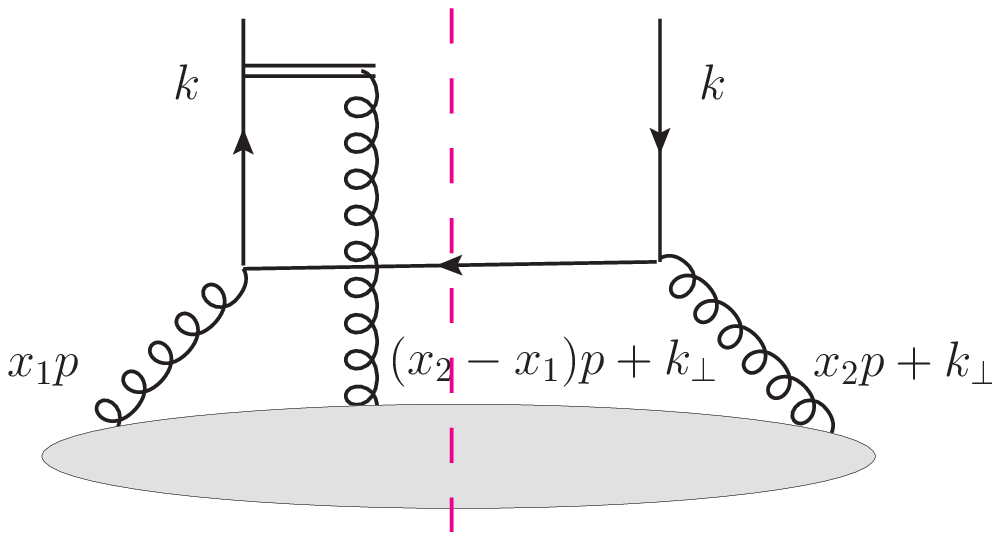, scale=0.5}
\caption{Feynman diagram for expansion of the quark Sivers function in terms of the three-gluon correlation function. The mirror diagram where middle gluon is to
the right of the unitary cut also contributes and is included in our final result.}
\label{sivers-3g}
\eef

In order to calculate the explicit $\pht$-dependence generated by the TMD factorization in Eq.~\eqref{eq:TMD-fac} (particularly those related to the quark Sivers function), following~\cite{Ji:2006br}, we let $k_\perp$ to be of the order of $\pht$ and the others ($\lambda_\perp$ and $p_\perp$) much smaller: $k_\perp \gg \Lambda_{\rm QCD} \sim \lambda_\perp, ~p_\perp$. In this case, we can neglect $\lambda_\perp$ and $p_\perp$ compared with $k_\perp$ in the delta function and obtain
\bea
\delta^2\left(z_h {\bf k_\perp} + {\bf p_\perp} + \boldsymbol{\lambda_\perp} - {\bf p_{h\perp}} \right) 
\stackrel{k_\perp\sim \pht}{ \xrightarrow{\hspace*{1cm}} } 
\delta^2\left(z_h {\bf k_\perp} - {\bf p_{h\perp}} \right).
\eea
At the same time, the integration over the other transverse momentum $p_\perp$ yields the ordinary collinear fragmentation function,
\bea
\int d^2{\bf p_\perp} D_{h/q}(z_h, p_\perp^2) = D_{h/q}(z_h),
\eea 
whereas the integration over $\boldsymbol{\lambda_\perp}$ leads to $\int d^2\boldsymbol{\lambda_\perp}\,S(\boldsymbol{\lambda_\perp}) = 1$, for details, see~\cite{Ji:2006br,Ji:2004wu,Ji:2004xq}. Finally, substituting the expansion of the quark Sivers function in terms of three-gluon correlation functions, Eq.~\eqref{eq:sivers-expand}, into the TMD factorization, Eq.~\eqref{eq:TMD-fac}, we find:
\bea
\left.\frac{d\Delta\sigma}{d\xb dy dz_h d^2\bpht}\right|_{\pht \ll Q} =& - z_h \sigma_0 \left(\epsilon^{\alpha\beta} s_\perp^\alpha \pht^\beta \right) \frac{1}{\left(\pht^2\right)^2} \sum_q e_q^2
\frac{\alpha_s}{2\pi^2} \int \frac{dz}{z} D_{h/q}(z) \delta(1-\hz)
\nnu
&\times \int \frac{dx}{x^2} P_{q\leftarrow g}(\hx) \left( \frac{1}{2}\right) \left[ O(x,x) + O(x,0) + N(x,x) - N(x,0)\right].
\eea
It is evident that the above result reproduces the transverse spin-dependent differential cross section in Eq.~\eqref{eq:twist-3}, the one derived from the collinear twist-3 factorization formalism. We have thus demonstrated the consistency between the collinear twist-3 factorization formalism and TMD factorization formalism for the twist-3 three-gluon correlation functions at moderate transverse momenta, $\Lambda_{\rm QCD} \ll \pht \ll Q$. 

In principle, to establish fully the connection between TMD and collinear twist-3 formalism at the first non-trivial order, one should also consider the situation where either $\lambda_\perp$ to be of the order $\pht$, or $p_\perp$ to be the order of $\pht$. Both situations were studied in~\cite{Ji:2006br}, and they establish the connection for the quark-gluon correlation function or the Qiu-Sterman function, which is different from what we have done here. 

Another result from our calculation above will be to obtain the QCD evolution equation of the Qiu-Sterman function $T_{q,F}(x,x)$, specifically the contribution from the three-gluon correlation functions. To achieve this, we start from Eq.~\eqref{eq:sivers-expand} and using the following identity on the left hand side \cite{Boer:2003cm,Kang:2011hk}, 
\bea
T_{q,F}(x_B, x_B, \mu_f^2) = - \frac{1}{M}\int^{\mu_f^2} d^2{\bf k_\perp} k_\perp^2 f_{1T}^{\perp q}(x_B, k_\perp^2)|_{\rm SIDIS},
\eea
we find (in the cut-off scheme)
\bea
T_{q,F}(x_B, x_B, \mu_f^2)  =  \int^{\mu_f^2} \frac{dk_\perp^2}{k_\perp^2}\, 
\frac{\alpha_s}{2\pi} \int_{x_B}^1 \frac{dx}{x^2}  P_{q\leftarrow g}(\hat x) \left(\frac{1}{2}\right)\left[O(x,x) + O(x,0) + N(x,x) - N(x,0)\right].
\eea
The evolution equation corresponding to the above expression is then
\bea
\frac{\partial}{\partial \ln\mu_f^2} T_{q,F}(x_B, x_B, \mu_f^2) = \frac{\alpha_s}{2\pi}\int_{x_B}^1 \frac{dx}{x^2}  P_{q\leftarrow g}(\hat x) \left(\frac{1}{2}\right)\left[O(x,x,\mu_f^2) + O(x,0,\mu_f^2) + N(x,x,\mu_f^2) - N(x,0,\mu_f^2)\right],
\label{eq:twist3-evo}
\eea
which is exactly the same as the one derived before from different approaches~\cite{Kang:2008ey,Braun:2009mi,Ma:2012xn}. In the next section, we will perform a complete NLO calculation for the $\pht$-weighted transverse spin-dependent cross section, and re-derive this evolution equation using dimensional regularization. 

%
%
\subsection{Coefficient functions in the TMD evolution formalism}

To study the QCD evolution of TMDs, one usually defines the TMDs in the Fourier conjugated 2-dimensional coordinate space - the so-called ``$b$-space''. For the quark Sivers function, the common definition in $b$-space is the following \cite{Kang:2011mr,Echevarria:2014xaa} \footnote{Note the proper defined TMDs depend on two additional scales, i.e., the factorization scale $\mu$ and another scale $\zeta$ associated with rapidity divergence. Here we suppress both dependences for simplicity.}
\bea
f_{1T}^{\perp q(\alpha)}(x_B, b) = \frac{1}{M} \int d^2{\bf k_\perp} e^{-i{\bf k_\perp}\cdot {\bf b}} k_\perp^\alpha f_{1T}^{\perp q}(x_B, k_\perp^2).
\eea
In the perturbative region $1/b \gg \Lambda_{\rm QCD}$, one can expand the above quark Sivers function $f_{1T}^{\perp q(\alpha)}(x_B, b) $ in terms of the corresponding collinear functions, i.e. the twist-3 Qiu-Sterman function $T_{q,F}(x_1, x_2)$ as well as the three-gluon correlation functions $O(x_1, x_2)$ and $N(x_1, x_2)$. If we collectively denote them as $f^{(3)}(x_1, x_2)$, we can write formally
\bea
f_{1T}^{\perp q(\alpha)}(x_B, b) =  \left(\frac{ib^\alpha}{2}\right) C_{q\leftarrow i}(\hat x_1, \hat x_2)\otimes f_{i}^{(3)}(x_1, x_2),
\eea
where $C_{q\leftarrow i}(\hat x_1, \hat x_2)$ is the coefficient function with $\hat x_{1,2} = x_B/x_{1,2}$. The precise meaning of the convolution~$\otimes$ will be defined below, where  the inclusion of the factor $\left(\frac{ib^\alpha}{2}\right)$ will also become clear. 

At leading order, one has \cite{Kang:2011mr,Aybat:2011ge,Echevarria:2014xaa}:
\bea
f_{1T}^{\perp q(\alpha)}(x_B, b) = \left(\frac{ib^\alpha}{2}\right) \int_{x_B}^1 \frac{dx}{x} \delta(1-\hx) T_{q,F}(x, x),
\eea
which tells us that the coefficient function $C_{q\leftarrow i}$ at leading order is given by
\bea
 C_{q\leftarrow i} = \delta_{qi} \delta(1-\hx).
\eea
Now, we will study the coefficient function $C_{q\leftarrow g}$ from the expansion of the quark Sivers function in terms of three-gluon correlation functions. To start, we redo the calculation which leads to Eq.~\eqref{eq:sivers-expand} in $n=4-2\epsilon$ dimensions, and obtain the following result:
\bea
\frac{1}{M}f_{1T}^{\perp q}(x_B, k_\perp^2) = & -\frac{\alpha_s}{2\pi^2}  \frac{\left(4\pi^2\mu^2\right)^\epsilon}{1-\epsilon} \frac{1}{\left(k_\perp^2\right)^2} \int_{x_B}^1 \frac{dx}{x^2}  
\Big\{
P_{q\leftarrow g}(\hat x) \left(\frac{1}{2}\right)\left[O(x,x) + O(x,0) + N(x,x) - N(x,0)\right]
\nnu
& -\frac{\epsilon}{4} \left[O(x,x)+N(x,x)\right]
-\epsilon \hx (1-\hx) \left[O(x,0) - N(x,0)\right]
\Big\},
\eea
where $\mu$ comes from the replacement $g\to g\mu^\epsilon$ in $n=4-2\epsilon$ dimensions. The factor $\frac{1}{1-\epsilon}$ on the right hand side comes from the following replacement in Eqs.~\eqref{eq:o} and \eqref{eq:n}:
\bea
\frac{1}{2} \to \frac{1}{2(1-\epsilon)}.
\label{half-issue}
\eea
Performing the Fourier transform 
\bea
f_{1T}^{\perp q(\alpha)}(x_B, b)  = \frac{1}{M} \int d^{2-2\epsilon}{\bf k_\perp} e^{-i{\bf k_\perp}\cdot {\bf b}} k_\perp^\alpha f_{1T}^{\perp q}(x_B, k_\perp^2),
\eea
we finally obtain the quark Sivers function in  $b$-space in terms of the three-gluon correlation functions:
\bea
f_{1T}^{\perp q(\alpha)}(x_B, b) = & \left(\frac{ib^\alpha}{2}\right) 
\Bigg\{
\frac{\alpha_s}{2\pi} \left(-\frac{1}{\hat \epsilon}\right) \int \frac{dx}{x^2} P_{q\leftarrow g}(\hat x) \left(\frac{1}{2}\right)\left[O(x,x) + O(x,0) + N(x,x) - N(x,0)\right]
\nnu
&+\frac{\alpha_s}{4\pi} \int \frac{dx}{x^2} \left[P_{q\leftarrow g}(\hx) \ln\left(\frac{c^2}{b^2\mu^2}\right)+\hx(1-\hx)\right] \left[O(x,x) + N(x,x)\right]
\nnu
&+\frac{\alpha_s}{4\pi}  \int \frac{dx}{x^2}  \left[P_{q\leftarrow g}(\hx) \ln\left(\frac{c^2}{b^2\mu^2}\right)-\frac{1}{2}\left(1-6\hx+6\hx^2\right)\right] \left[O(x,0) - N(x,0)\right]
\Bigg\},
\label{eq:sivers-b}
\eea
where $1/\hat \epsilon=1/\epsilon-\gamma_E+\ln 4\pi$ and $c=2e^{-\gamma_E}$. To arrive at the above result, we have used the following identity:
\bea
\int \frac{d^n{\bf k_\perp}}{(2\pi)^n}\frac{1}{\left(k_\perp^2\right)^m}e^{-ik_\perp \cdot b}
=\frac{1}{(4\pi)^{n/2}}\frac{\Gamma\left(\frac{n}{2}-m\right)}{\Gamma(m)}\left(\frac{b^2}{4}\right)^{m-n/2}.
\eea
The terms with $k_\perp^\alpha$ in the integrand can be derived by taking derivative with respect to $b^\alpha$ from the above formula.

It is instructive to realize that the term $\propto -1/\hat\epsilon$ in Eq.~\eqref{eq:sivers-b} is simply the $\mathcal O(\alpha_s)$ correction to the Qiu-Sterman function $T_{q,F}(x,x)$ (recall the evolution equation in Eq.~\eqref{eq:twist3-evo}), and  should thus  be subtracted in the definition of the perturbative coefficient functions~\cite{Collins:2011zzd,Bacchetta:2013pqa}. If we write the coefficient functions as follows:
\bea
f_{1T}^{\perp q(\alpha)}(x_B, b) = \left(\frac{ib^\alpha}{2}\right) \int_{x_B}^1\frac{dx}{x^2} \Big\{C_{q\leftarrow g, 1}(\hat x) \big[O(x, x)+N(x,x)\big]+ C_{q\leftarrow g, 2}(\hat x) \big[O(x,0) - N(x,0)\big] \Big\},
\eea
we then have 
\bea
C_{q\leftarrow g,1}(\hx) &= \frac{\alpha_s}{4\pi} \left[P_{q\leftarrow g}(\hx) \ln\left(\frac{c^2}{b^2\mu^2}\right)+\hx(1-\hx)\right],
\\
C_{q\leftarrow g,2}(\hx) &= \frac{\alpha_s}{4\pi} \left[P_{q\leftarrow g}(\hx) \ln\left(\frac{c^2}{b^2\mu^2}\right)-\frac{1}{2}\left(1-6\hx+6\hx^2\right)\right].
\eea 
To summarize, we have derived the coefficient functions $C_{q\leftarrow g}$ when expanding the {\it unsubtracted} quark Sivers function in terms of the three-gluon correlation functions. However, it is important to point out that such coefficient functions will be exactly {\it the same} even if one uses the new properly defined TMDs in \cite{Collins:2011zzd} and/or \cite{Echevarria:2012js,Echevarria:2012pw} \footnote{This fact only applies to the off-diagonal coefficients $C_{q\leftarrow g}$. For the diagonal ones $C_{q\leftarrow q}$, one has to include the contributions from the additional soft factors in the definition of TMDs.}. This is because at order ${\mathcal O}(\alpha_s)$ there is no contribution from soft factor subtraction \cite{Collins:2011zzd, Bacchetta:2013pqa}. Thus, one can use the coefficient functions derived above in the standard TMD evolution formalism \cite{Aybat:2011ge,Echevarria:2014xaa}.

%
%
\section{Transverse momentum weighted spin-dependent cross section}
In this section we study the transverse momentum-weighted transverse spin-dependent cross section at next-to-leading order. Again we focus on the {\it light} hadron production, as opposed to the massive open charm production in~\cite{Beppu:2010qn}. Such a transverse momentum-weighted transverse spin-dependent cross section is defined as~\cite{Kang:2012ns}:
\bea
\frac{d\langle \pht \Delta\sigma\rangle}{dx_B dy dz_h}
\equiv
\int d^2{\bpht} \epsilon^{\alpha\beta} s_{\perp}^\alpha \pht^{\beta} \frac{d \Delta\sigma}{dx_B dy dz_hd^2\bpht}. 
\label{pht-weight}
\eea
The leading order result is proportional to the twist-3 quark-gluon correlation function $T_{q,F}(x_1,x_2)$ and is given by~\cite{Kang:2012ns}
\bea
\frac{d\langle \pht \Delta\sigma\rangle}{dx_B dy dz_h} = -\frac{z_h\sigma_0}{2} \sum_q e_q^2
\int \frac{dx}{x} \frac{dz}{z} T_{q,F}(x,x) D_{h/q}(z) \delta(1-\hat x)\delta(1-\hat z).
\label{lo-res}
\eea
Since we will compute such a $\pht$-weighted cross section at NLO, which contains divergences, we will present all  results in $n=4-2\epsilon$ dimensions, in which $\sigma_0$ in Eq.~\eqref{lo-res} is given by
\bea
\sigma_0 = \frac{2 \pi \alpha_{em}^2}{Q^2}\frac{1+(1-y)^2}{y}(1-\epsilon).
\eea

The NLO correction from the photon-quark scattering channel $\gamma^*+q\to q+g$ has already been computed in Ref.~\cite{Kang:2012ns}, from which one derives the QCD evolution equation (factorization scale $\mu_f$-dependence) of the Qiu-Sterman function $T_{q, F}(x, x, \mu_f^2)$, in particular the contribution from itself. In this section, we study the photon-gluon scattering channel $\gamma^*+g\to q+\bar q$. This will allow us to study the contribution to the QCD evolution  of $T_{q, F}(x, x, \mu_f^2)$ from the twist-3 three-gluon correlation functions.

The relevant Feynman diagrams are given in Fig.~\ref{fig:soft-pole}. The calculation is almost the same as the one presented in Sec.~\ref{cross section}, i.e. it starts from the ``master formula'' in Eq.~\eqref{eq:w1} to derive the final result in Eq.~\eqref{result-4}. The only differences are as following: first, one has to perform all the calculations in $n=4-2\epsilon$ dimensions, while Eq.~\eqref{result-4} is the result in the usual 4 dimensions; second, one has to perform the $\pht$-weight as specified in the definition Eq.~\eqref{pht-weight}. There are a couple of places where one has to exercise extra care to ensure that the correct finite NLO corrections are obtained: first, for both Eqs.~\eqref{eq:o} and \eqref{eq:n}, one has to make the replacement as specified in Eq.~\eqref{half-issue}; second, one will encounter the following replacement in the calculations
\bea
\pht^{\beta} \pht^{\sigma}\to \frac{1}{2(1-\epsilon)} \pht^2 g_\perp^{\beta\sigma}.
\eea
Finally, the $\pht$-weighted spin-dependent differential cross section can be written as follows:
\bea
\frac{d\langle\pht\Delta\sigma\rangle}{d\xb dy dz_h} =& z_h \sigma_0 \sum_q e_q^2 \left(\frac{1}{4}\right) \frac{\alpha_s}{2\pi}\int\frac{dx}{x}\frac{dz}{z} D_{h/q}(z) \left( \frac{4\pi\mu^2}{Q^2}\right)^\epsilon \frac{1}{\Gamma(1-\epsilon)}     
\frac{1}{2(1-\epsilon)^2} 
\hz^{-\epsilon}(1-\hz)^{1-\epsilon}\hx^{\epsilon-1}(1-\hx)^{1-\epsilon}
\nnu
&\times \left\{\left[\left(\frac{d O(x,x)}{d x}-\frac{2 O(x,x)}{x}\right)H_1+\left(\frac{d O(x,0)}{d x}-\frac{2 O(x,0)}{x}\right)H_2+\frac{O(x,x)}{x}H_3+\frac{O(x,0)}{x}H_4\right]  \right. 
\nnu
&+\left.\left[\left(\frac{d N(x,x)}{d x}-\frac{2 N(x,x)}{x}\right)H_1-\left(\frac{d N(x,0)}{d x}-\frac{2 N(x,0)}{x}\right)H_2+\frac{N(x,x)}{x}H_3-\frac{N(x,0)}{x}H_4\right]\right\},
\label{final-0}
\eea
where the hard-part functions are given by
\bea
H_1&=\frac{\hx\left[2\hx^2-2\hx+(1-2\hz+2\hz^2)-\epsilon\right]}{\hz^2(1-\hz)^2},
\\
H_2&=\frac{\hx\left[4\hx^2-4\hx+(1-2\hz+2\hz^2)-\epsilon(2\hx-1)^2\right]}{\hz^2(1-\hz)^2(1-\epsilon)},
\\
H_3&=\frac{2\hx^2\left(1-2\hx\right)}{\hz^2(1-\hz)^2},
\\
H_4&=\frac{2\hx^2\left(1-4\hx+2\epsilon\right)}{\hz^2(1-\hz)^2}.
\eea

The next step will be to perform the $\epsilon$-expansion for our result in Eq.~\eqref{final-0} and isolate the divergence part and the finite NLO correction contributions. To simplify our notation, let us define
\bea
I=\hz^{-\epsilon}(1-\hz)^{1-\epsilon}\hx^{\epsilon-1}(1-\hx)^{1-\epsilon}.
\eea
We carry out the $\epsilon$-expansion for the products $I\times (H_1, H_2, H_3, H_4)$, which have the following results:
\bea
I\times H_1=&-\frac{1}{\epsilon}(1-\hx)(2\hx^2-2\hx+1)\delta(1-\hz) - \hat H_1,
\label{eq:full-H1}
\\
I\times H_2=&-\frac{1}{\epsilon}(1-\hx)(4\hx^2-4\hx+1)\delta(1-\hz) - \hat H_2,
\label{eq:full-H2}
\\
I\times H_3=&-\frac{1}{\epsilon}(1-\hx)2\hx(1-2\hx)\delta(1-\hz) - \hat H_3,
\label{eq:full-H3}
\\
I\times H_4=&-\frac{1}{\epsilon}(1-\hx)2\hx(1-4\hx)\delta(1-\hz) - \hat H_4.
\label{eq:full-H4}
\eea
Here the {\it finite} hard-part functions $\hat H_{i=1,2,3,4}$ are given by
\bea
\hat H_1=&\delta(1-\hz)(1-\hx)\left[(2\hx^2-2\hx+1)\left(\ln\frac{\hx}{1-\hx}+2\right)-1\right] 
-\frac{(1-\hx)(2\hx^2-2\hx+1-2\hz+2\hz^2)}{\hz^2(1-\hz)_+},
\label{eq:finite-H1}
\\
\hat H_2=&\delta(1-\hz)(1-\hx)(1-2\hx)^2\left(\ln\frac{\hx}{1-\hx}+3\right) 
-\frac{(1-\hx)(4\hx^2-4\hx+1-2\hz+2\hz^2)}{\hz^2(1-\hz)_+},
\label{eq:finite-H2}
\\
\hat H_3=&\delta(1-\hz)(1-\hx)2\hx(1-2\hx)\left(\ln\frac{\hx}{1-\hx}+2\right) 
-\frac{(1-\hx)2\hx(1-2\hx)}{\hz^2(1-\hz)_+},
\label{eq:finite-H3}
\\
\hat H_4=&\delta(1-\hz)(1-\hx)2\hx\left[(1-4\hx)\left(\ln\frac{\hx}{1-\hx}+2\right)+2\right] 
-\frac{(1-\hx)2\hx(1-4\hx)}{\hz^2(1-\hz)_+}.
\label{eq:finite-H4}
\eea

Let us now concentrate on the divergent $1/\epsilon$ terms as given in Eqs.~\eqref{eq:full-H1}, \eqref{eq:full-H2}, \eqref{eq:full-H3}, and \eqref{eq:full-H4}, which are all proportional to $\delta(1-\hz)$. We collect these terms, perform integration by parts to convert all the derivative terms to non-derivative terms, and find the following expression:
\bea
\frac{d\langle\pht\Delta\sigma\rangle}{d\xb dy dz_h} =& -\frac{z_h\sigma_0}{2} \sum_q e_q^2\int \frac{dz}{z}
D_{h/q}(z) \delta(1-\hz) \left(-\frac{1}{\hat \epsilon} + \ln\left(\frac{Q^2}{\mu^2}\right)\right)
\nnu
&\times
\frac{\alpha_s}{2\pi} \int \frac{dx}{x^2} P_{q\leftarrow g}(\hx) \left(\frac{1}{2}\right) \left[ O(x,x) + O(x,0) + N(x,x) - N(x,0)\right] + \cdots,
\label{eq:divergent}
\eea
where the ``$\cdots$'' represents the finite NLO corrections and will be given below in Eq.~\eqref{eq:final-result}.  By comparing Eq.~\eqref{eq:divergent} to the LO result in Eq.~\eqref{lo-res}, one realizes that the divergent part should be the collinear QCD correction to the LO bare Qiu-Sterman function $T_{q,F}^{(0)}(x_B, x_B)$ that is absorbed into the definition of the renormalized $T_{q,F}(x_B, x_B)$ as follows:
\bea
T_{q,F}(x_B, x_B, \mu_f^2) =& T_{q,F}^{(0)}(x_B, x_B) +  \left(-\frac{1}{\hat \epsilon}+\ln\left(\frac{\mu_f^2}{\mu^2}\right)\right) \frac{\alpha_s}{2\pi}
\int_{x_B}^1 \frac{dx}{x^2}  
\nnu
&\times P_{q\leftarrow g}(\hat x) \left(\frac{1}{2}\right)
\left[O(x,x) + O(x,0) + N(x,x) - N(x,0)\right],
\label{eq:TF-NLO}
\eea
where we have adopted $\overline{\rm MS}$-scheme and $\mu_f$ is the factorization scale. From Eq.~\eqref{eq:TF-NLO}, one can obtain the evolution equation for the Qiu-Sterman function (the off-diagonal contribution from three-gluon correlation functions):
\bea
\frac{\partial}{\partial \ln\mu_f^2} T_{q,F}(x_B, x_B, \mu_f^2) = \frac{\alpha_s}{2\pi}\int_{x_B}^1 \frac{dx}{x^2}  P_{q\leftarrow g}(\hat x) \left(\frac{1}{2}\right)\left[O(x,x,\mu_f^2) + O(x,0,\mu_f^2) + N(x,x,\mu_f^2) - N(x,0,\mu_f^2)\right].
\eea 
This result confirms our result derived above in the cut-off scheme, Eq.~\eqref{eq:twist3-evo}, and also agrees with the earlier findings~\cite{Kang:2008ey,Braun:2009mi,Ma:2012xn}.

After the $\overline{\rm MS}$ subtraction of the collinear divergence into the Qiu-Sterman function $T_{q,F}(x_B, x_B, \mu_f^2)$, we obtain the NLO corrections for the three-gluon correlation functions to the $\pht$-weighted transverse spin-dependent differential cross section:
\bea
\frac{d\langle\pht\Delta\sigma\rangle}{d\xb dy dz_h} =& - \frac{z_h \sigma_0}{2} \frac{\alpha_s}{2\pi} \sum_q e_q^2  \int_{x_B}^1\frac{dx}{x}  \int_{z_h}^1 \frac{dz}{z} D_{h/q}(z) \Bigg\{\delta(1-\hz) \ln\left(\frac{Q^2}{\mu_f^2}\right) P_{q\leftarrow g}(\hat x)
\nnu
&\times \left(\frac{1}{2x}\right)
\left[O(x,x,\mu_f^2) + O(x,0,\mu_f^2) + N(x,x,\mu_f^2) - N(x,0,\mu_f^2)\right]
\nnu
&+\left(\frac{1}{4}\right) 
\left[\left(\frac{d O(x,x, \mu_f^2)}{d x}-\frac{2 O(x,x,\mu_f^2)}{x}\right) \hat H_1+\left(\frac{d O(x,0,\mu_f^2)}{d x}-\frac{2 O(x,0,\mu_f^2)}{x}\right)\hat H_2\right.
\nnu
&\left.+\frac{O(x,x,\mu_f^2)}{x}\hat H_3+\frac{O(x,0,\mu_f^2)}{x}\hat H_4\right] 
+\left(\frac{1}{4}\right)  \left[\left(\frac{d N(x,x,\mu_f^2)}{d x}-\frac{2 N(x,x,\mu_f^2)}{x}\right)\hat H_1\right.
\nnu
&-\left(\frac{d N(x,0,\mu_f^2)}{d x}-\frac{2 N(x,0,\mu_f^2)}{x}\right)\hat H_2
\left.+\frac{N(x,x,\mu_f^2)}{x}\hat H_3-\frac{N(x,0,\mu_f^2)}{x}\hat H_4\right]
\Bigg\},
\label{eq:final-result}
\eea
where the finite hard-part functions $\hat H_{i=1,2,3,4}$ are given in Eqs.~\eqref{eq:finite-H1}, \eqref{eq:finite-H2}, \eqref{eq:finite-H3}, and \eqref{eq:finite-H4}, respectively. The result follows the standard form expected from collinear factorization, i.e. the logarithm containing the factorization scale together with the splitting function (the first line above) determines the evolution of the twist-3 Qiu-Sterman function in terms of the three-gluon correlation functions. 
 
%
%
\section{Conclusions}
In this paper we calculated the contribution of the three-gluon correlation functions to the Sivers asymmetry for semi-inclusive hadron production in deep inelastic scattering. Within the twist-3 collinear factorization formalism, we first studied the unweighted spin-dependent differential cross section. We then demonstrated that the result derived in such a framework is consistent with the one obtained from the transverse momentum dependent factorization at moderate hadron transverse momenta, $\Lambda_{\rm QCD}\ll \pht \ll Q$. This extends the unification of the two mechanisms to include the case of three-gluon correlation functions. In the process of this demonstration, we also derived the ${\mathcal O}(\alpha_s)$ expansion of the quark Sivers function in terms of the three-gluon correlation functions, the so-called {\it off-diagonal} piece. One might also use our approach to study the expansion of the gluon Sivers function in terms of the three-gluon correlation functions, which is usually referred to as the {\it diagonal} contributions. We leave such a study for future work. From the expansion expression, we identified the so-called coefficient functions that are used in the usual TMD evolution formalism.  We further calculated the next-to-leading order perturbative QCD corrections to the transverse-momentum-weighted spin-dependent differential cross section, from which we identified the off-diagonal contribution from the three-gluon correlation functions to the QCD collinear evolution of the twist-3 Qiu-Sterman function. We found that our evolution equation agrees with those derived previously from different approaches. 

\section*{Acknowledgments}
This material is based upon work supported by the U.S. Department of Energy, Office of Science, Office of Nuclear Physics under contracts DE-AC05-06OR23177 (L.D., A.P.) and DE-AC52-06NA25396 (Z.K., I.V.), and in part by the LDRD program at LANL.

\end{document}